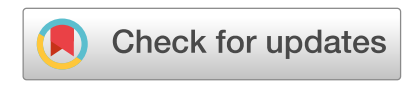

OPEN

# Secure multiparty quantum key agreement against collusive attacks

Hussein Abulkasim[1,2]✉, Atefeh Mashatan[1] & Shohini Ghose[3,4]

**Quantum key agreement enables remote participants to fairly establish a secure shared key based on their private inputs. In the circular-type multiparty quantum key agreement mode, two or more malicious participants can collude together to steal private inputs of honest participants or to generate the final key alone. In this work, we focus on a powerful collusive attack strategy in which two or more malicious participants in particular positions, can learn sensitive information or generate the final key alone without revealing their malicious behaviour. Many of the current circular-type multiparty quantum key agreement protocols are not secure against this collusive attack strategy. As an example, we analyze the security of a recently proposed multiparty key agreement protocol to show the vulnerability of existing circular-type multiparty quantum key agreement protocols against this collusive attack. Moreover, we design a general secure multiparty key agreement model that would remove this vulnerability from such circular-type key agreement protocols and describe the necessary steps to implement this model. The proposed model is general and does not depend on the specific physical implementation of the quantum key agreement.**

The concept of the key agreement was first presented by Diffie–Hellman in 1976[1]. It describes how two remote users can fairly generate a secured shared key based on their private inputs. In 1982, Ingemarsson et al.[2] extended the two-party key agreement protocol to a multiparty or group key agreement protocol. After that, several multiparty key agreement schemes have been published[3]. However, future quantum computers with sufficient power could threaten the security of many of the current public-key cryptosystems whose security mainly relies on unproven mathematical assumptions[4,5]. For that reason, quantum applications in cryptography have attracted the attention of a lot of scientists and researchers to suggest and develop information-theoretically unconditional secure cryptosystems. One of the most common quantum cryptographic applications is quantum key distribution (QKD)[6], in which remote participants can establish a shared random key securely even in the presence of an attacker with unlimited classical or quantum computing power. Subsequently, several quantum cryptographic applications have been introduced to solve various classical security issues[7–16].

Recently, quantum key agreement (QKA) has attracted the attention of a lot of researchers[17]. QKA ensures fairness between the involved participants to generate a shared secure key based on their private inputs. Using the quantum teleportation protocol, Zhou et al.[17], in 2004, presented the first two-party QKA scheme. In 2013, the two-party QKA was extended to multiparty QKA protocols[18]. Subsequently, several multiparty QKA protocols have been presented[19–22]. In general, as noted in[18], there are three types of MQKA protocols; (1) the first type is the tree-type in which every party sends their secret data through independent quantum channels to all other parties[23]; (2) the second type is the complete-graph-type in which every participant sends a sequence of qubits to each of the others parties to encode her or his secret information, (3) while in the third type that is the circle-type (sometimes called travelling-mode)[8,24], every party generates a random sequence of qubits and sends this sequence to another party who applies an encoding process producing a new evolved sequence of qubits and sends the new sequence to the next party; this process continues over all parties until the evolved sequence reaches the party who generates the first sequence. Compared to the other QKA types, the circle-type is more efficient and more easily achieves the property of fairness. For that reason, the QKA circle-type has been intensively investigated.

In 2016, Liu et al.[18] pointed out that all existing circle-type multiparty quantum key agreement (CT-MQKA) protocols are vulnerable to collusive attack, and asked a challenging question about the possibility of designing a secure CT-MQKA protocol. In response to this question, several CT-MQKA protocols have been proposed to avoid collusive attacks. However, in this work, we show that many of the existing CT-MQKA protocols are also not secure against a collusive attack. We study, as an example, the security of Sun et al.'s[19] MQKA protocol (named SCWZ protocol hereafter) to show the vulnerability of the existing CT-MQKA protocols to collusive attacks. Furthermore, we design a general secure model for CT-MQKA protocols and propose the necessary steps for this model.

[1]Ted Rogers School of Information Technology Management, Ryerson University, Toronto, Canada. [2]Faculty of Science, New Valley University, El-Kharga, Egypt. [3]Department of Physics and Computer Science, Wilfrid Laurier University, Waterloo, Canada. [4]Institute for Quantum Computing, University of Waterloo, Waterloo, Canada. ✉email: abulkasim@ryerson.ca

## The insecurity of existing CT-MQKA protocols

In this section, we show that many of the recently published works in CT-MQKA are not secure against collusive attacks[19–21,25,26]. In general, there are two main collusive attack strategies, which could be applied to the CT-MQKA protocols:

**The first collusive attack strategy.** The first collusive attack strategy has been pointed out in[18,19]. Any two dishonest participants $P_i$ and $P_j$ (where $i > j$; $i, j \in \{1, 2, \ldots, n\}$ and $n$ is the number of participants) in particular positions in the circle-type protocols can control the final key if their particular positions meet the following two conditions:

$$i - j = \frac{n}{2} \text{ when } n \text{ is even,} \tag{1}$$

$$i - j = \frac{n+1}{2} \text{ or } \frac{n-1}{2} \text{ when } n \text{ is odd.} \tag{2}$$

**The second collusive attack strategy.** The second collusive attack strategy can be described as follows. In the CT-MQKA schemes, any two dishonest participants $P_i$ and $P_j$ can steal the private inputs of an honest participant $P_k$ ($i, j, k \in \{1, 2, \ldots, n\}$) without being detected, if their positions meet one of the two following conditions:

$$i - j = 2; \text{ then k} = i - 1; \tag{3}$$

$$j - i = 2; \text{ then k} = j - 1. \tag{4}$$

Note, in our previous work[8], we mentioned that two malicious users may try to deduce the private information of an honest one. However, in this work, we formulate and describe the general situation in which two dishonest participants can steal the private information of the honest ones as indicated in Eqs. (3) and (4).

**Review of SCWZ's protocol.** In SCWZ's protocol[19], there are $n$ participants and each participant $P_i$ ($i = 1, 2, \ldots, n$) has an $m - bit$ key ($K_i$). The participants want to generate a shared secret key $K$ fairly, where $K = K_1 \oplus K_2 \oplus \cdots \oplus K_n$. The steps of the SCWZ's protocol can be described as follows.

1) Preparation phase. The server generates $n$ sequences of random single-photons. Each sequence $S_i$ ($i = 1, 2, \ldots, n$) contains $m$ single-photons and each photon is selected randomly from the four states $\{|+\rangle, |-\rangle, |0\rangle, |1\rangle\}$, where $|\pm\rangle = \frac{1}{\sqrt{2}}(|0\rangle \pm |1\rangle)$. The server also generates $n$ sequences of random single photons (called $C_i$), which are used as decoy photons to check the existence of eavesdroppers. Each single decoy photon is randomly selected from the states $\{|+\rangle, |-\rangle, |+y\rangle, |-y\rangle\}$, where $|\pm y\rangle = \frac{1}{\sqrt{2}}(|0\rangle \pm i|1\rangle)$. The server then randomly inserts and distributes the single-photons of $C_i$ into $S_i$ getting a new sequence $S_i^{'}$, and sends the new sequence ($S_i$) to $P_i$.
2) Detection phase. Upon receiving $S_i^{'}$, each participant sends an acknowledgment to the server. Then the server announces the positions of $C_i$ and their measurement bases. Each $P_i$ measures $C_i$ based on the corresponding measurement bases and stores the results. $P_i$ then randomly announces half of the measurement results of $C_i$; the server, in turn, announces the initial states of the second half of $C_i$. Then both the server and $P_i$ collaborate to compute the error rate. They end the protocol if the error rate higher than a predefined value. Otherwise, they continue with the protocol.
3) After $P_i$ gets the secure sequence $S_i$, each participant performs the next sub-steps:
   A. Encoding phase. $P_i$ encodes secret information ($K_i$) onto $S_i$ by applying the unitary operation $U = |0\rangle\langle 1| - |1\rangle\langle 0|$ when the classical bit of the secret $K_i$ is 1, and the unitary operation $I = |0\rangle\langle 0| + |1\rangle\langle 1|$ when the classical bit $K_i$ is 0. $P_i$ then reorders the decoy states that were prepared and inserted by the server in Step (1) and reinserts them in random positions into the encoded sequence obtaining a new sequence ($S_i^{i+1}$), and sends $S_i^{i+1}$ to $P_{i+1}$.
   B. Eavesdropping check phase. Upon receiving $S_i^{i+1}$, $P_{i+1}$ and $P_i$ check the security of the transmission by performing the same process indicated in step (2) between the server and $P_i$.
   C. Encoding phase. After checking the security of transmission, $P_{i+1}$ encodes secret information ($K_{i+1}$) onto $S_i$ following the same rules as in step (A). $P_{i+1}$ then reorders the decoy states and reinserts them in random positions into the encoded sequence obtaining a new sequence ($S_i^{i+2}$), and sends $S_i^{i+2}$ to $P_{i+2}$.
   D. Similarly, the rest of the participants ($P_{i+2}, P_{i+3}, \ldots, P_{i-2}$) perform the *Eavesdropping check phase* and the *Encoding phase* indicated in steps (B) and (C).

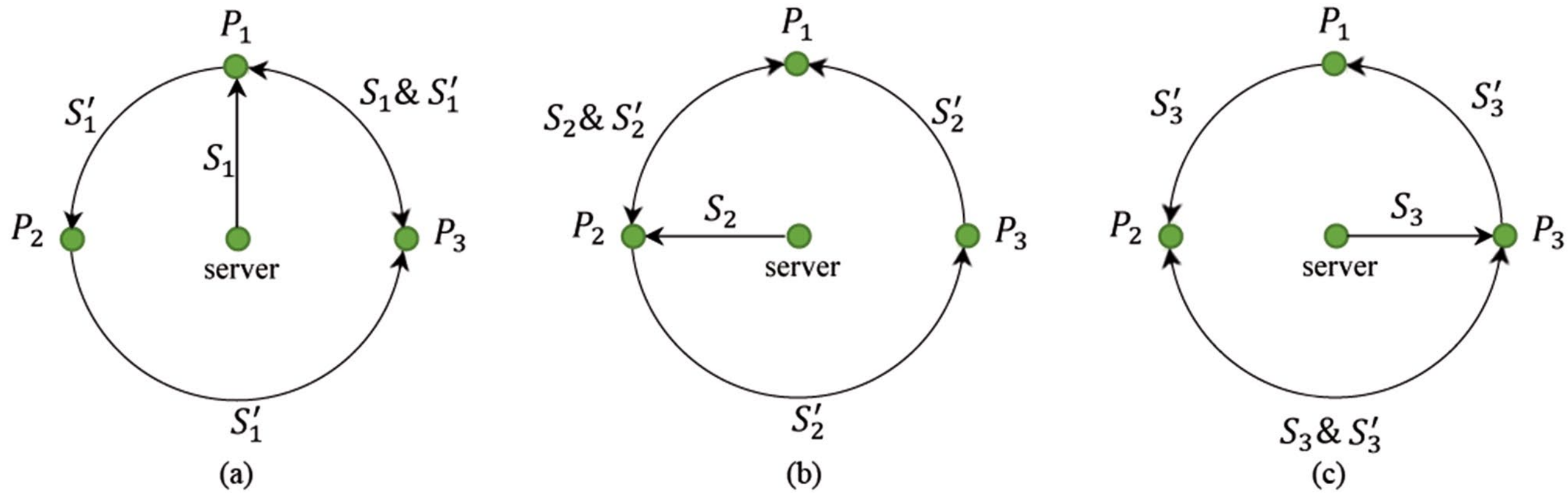

**Figure 1.** An example of a three-party QKA protocol. Any two dishonest participants in particular positions can steal the private input of an honest participant.

E. Upon receiving $S_i^{i-2}$, $P_{i-1}$ and $P_{i-2}$ check the security of transmission. If the quantum channel between $P_{i-1}$ and $P_{i-2}$ is secure, $P_{i-2}$ discards the decoy photons to get $S_i$, and informs the server of this fact.

4) When all the $P_{i-1}$ receive $S_i$, they send an acknowledgment to the server, and the server announces the measurement bases of $S_i$ to all the $P_{i-1}$. After that, each $P_{i-1}$ uses the corresponding measurement bases to measure $S_i$ obtaining $K_i^{'}$, where $K_i' = K_i \oplus K_{i+1} \oplus \cdots \oplus K_{i-2}$. Finally, $P_{i-1}$ can recover the final shared secret key $K = K_i^{'} \oplus K_{i-1}$.

**The collusive attack against CT-MQKA protocols.** In this section, we show that the SCWZ's protocol, as an example of CT-MQKA protocols, is insecure against a collusive attack. Although the authors of SCWZ's protocol have presented a security analysis to prove the security of their protocol against the first model of the collusive attack, their protocol is not secure against the second security model of collusive attack. That is to say, any two dishonest participants $P_i$ and $P_j$ in particular positions meeting the conditions in (3) and (4) can easily steal the private key of honest participants ($P_k$).

Without loss of generality, assume we have three participants $P_1$, $P_2$, and $P_3$ and they have three private keys, e.g., $K_1 = 1000$, $K_2 = 0101$, and $K_3 = 1001$, respectively. And the three participants intend to share a secret key ($K$), here $K = K_1 \oplus K_2 \oplus K_3 = 0100$. We also assume that $P_1$ and $P_3$ are two dishonest participants and they need to steal the private key of the honest one ($P_2$); hence they can deduce the final key without being caught.

In SCWZ's protocol, the server generates three random sequences, e.g., $S_1 = \{|+\rangle, |0\rangle, |1\rangle, |-\rangle\}$, $S_2 = \{|0\rangle, |1\rangle, |0\rangle, |1\rangle\}$, and $S_3 = \{|0\rangle, |+\rangle, |-\rangle, |1\rangle\}$ each one consists of four single-photons. Also, the server generates three random sequences $C_1$, $C_2$, and $C_3$ each one consists of four decoy single-photon states. Then the server randomly inserts the decoy state $C_1$ ($C_2$/$C_3$) into $S_1 = \{|+\rangle, |0\rangle, |1\rangle, |-\rangle\}$ ($S_2 = \{|0\rangle, |1\rangle, |0\rangle, |1\rangle\}$ /$S_3 = \{|0\rangle, |+\rangle, |-\rangle, |1\rangle\}$) and sends it to $P_1$ ($P_2$/$P_3$). After checking the security of the transmission, each participant discards the decoys and encodes their private information based on the encoding rule mentioned in Step 3.A. Subsequently, each participant sends the sequence in a circle to the other participants to encode their private inputs until the sequence is returned to the participant.

For simplicity, we show here the circle of $S_1$ (Fig. 1a) which will be used by the participant $P_1$ to get the final key ($K$). First, $P_1$ receives $S_1 = \{|+\rangle, |0\rangle, |1\rangle, |-\rangle\}$ from the server to encode her or his information and get the final key. Second, $P_1$ encodes a private input, i.e., $K_1 = 1000$ into $S_1$ getting the new sequence $S_1 = \{U|+\rangle, I|0\rangle, I|1\rangle, I|-\rangle\}$. Third, $P_1$ inserts some decoy photons into $S_1$ and sends it to the dishonest $P_3$ instead of sending it to $P_2$. After checking the security of the transmission, $P_3$ discards the decoy states and gets $S_1 = \{U|+\rangle, I|0\rangle, I|1\rangle, I|-\rangle\}$. At the same time, the dishonest $P_1$ generates a counterfeit sequence, e.g., $S_1^{'} = \{|0\rangle, |0\rangle, |-\rangle, |+\rangle\}$ with decoy states and sends it to both $P_2$ and $P_3$. $P_1$ only tells $P_3$ that the sequence $S_1^{'}$ is the counterfeit one. Since the honest participant ($P_2$) does not have $K_1 = 1000$ and does not knows $S_1 = \{|+\rangle, |0\rangle, |1\rangle, |-\rangle\}$, she or he cannot know what the received new sequence looks like (i.e., $S_1 = \{U|+\rangle, I|0\rangle, I|1\rangle, I|-\rangle\}$).

Obviously, $P_2$ cannot distinguish between the genuine sequences and the counterfeit ones. So, $P_2$ encodes the private data, i.e., $K_2 = 0101$ into $S_1^{'}$ getting $S_1^{'} = \{I|0\rangle, U|0\rangle, I|-\rangle, U|+\rangle\}$ and sends $S_1^{'}$ with decoy states to $P_3$. After checking the security of the transmission, $P_3$ discards the decoy qubits and gets $S_1^{'}$. $P_3$ then requests the corresponding measurement bases of $S_1^{'}$ from $P_1$ to get $K_2 = 0101$. Based on her or his private key, i.e., $K_3 = 1001$ and the private key of $P_2$, $P_3$ applies the corresponding unitary operations to the genuine sequence $S_1 = \{U|+\rangle, I|0\rangle, I|1\rangle, I|-\rangle\}$ getting $S_1 = \{U(I(U|+\rangle)), I(U(I|0\rangle)), I(I(I|1\rangle)), U(U(I|-\rangle))\}$ and sends it to $P_1$. Then the participants announce to the server that the quantum channels are secure. Finally, the server announces the measurement bases of $S_1$ to $P_1$ thus enabling $P_1$ to get K.

Similarly, if $P_2$ and $P_3$ ($P_2$ and $P_1$) are the dishonest participants they can steal the private key of the honest participant $P_1$ ($P_3$) in the circle while sending $S_2$ ($S_3$), as shown in Fig. 1b (Fig. 1c). By applying the same attack strategy, most of the existing CR-MQKA protocols[19–21,25,26] are vulnerable to this collusive attack.

| Unitary operations/quantum states | $\|0\rangle$ | $\|1\rangle$ | $\|+\rangle$ | $\|-\rangle$ |
|---|---|---|---|---|
| $0 \Rightarrow I$ | $\|0\rangle$ | $\|1\rangle$ | $\|+\rangle$ | $\|-\rangle$ |
| $1 \Rightarrow U$ | $-\|1\rangle$ | $\|0\rangle$ | $\|-\rangle$ | $-\|+\rangle$ |

**Table 1.** The encoding rules. The unitary operation $I$ represents 0 and the unitary operation $U$ represents 1.

## The proposed secure CT-MQKA protocol

In this section, we give a general secure model of CT-MQKA against the collusive attack described above. Whereas our protocol can be implemented with photons, we describe it in more general terms here. The idea of adopting a semi-honest client–server model (or a third party) has been adopted in many previous QKA protocols to ensure the security of communication[19,22,27–29]. Suppose we have $n$ participants who want to generate a shared secret key $K$ fairly, where $K = K_1 \oplus K_2 \oplus \cdots \oplus K_n$ with length $m$. Every participant ($P_i$) selects a private random classical key ($K_i^{'}$), where $\left|K_i^{'}\right| = |K_i| + nl$. Here, $l$ is the number of decoy states used for checking the security of a quantum channel, and $i = 1, 2, \ldots n$.

The general steps of this secure CT-MQKA model can be described as follows:

**Step (1)** The server generates $n$ sequences $S_i$ ($i = 1, 2, \ldots, n$), with each sequence containing $m + nl$ single qubits. The server records the position of every single qubit. Every qubit is selected randomly from the four quantum states $\left\{|+\rangle = \frac{1}{\sqrt{2}}(|0\rangle + |1\rangle), |-\rangle = \frac{1}{\sqrt{2}}(|0\rangle - |1\rangle), |0\rangle, |1\rangle\right\}$.

**Step (2)** The server also generates additional $n$ sequences of random single qubits (called $C_i$), which are used as decoy states to check the existence of eavesdroppers. Every single decoy qubit is randomly selected from the four quantum states $\{|+\rangle, |-\rangle, |0\rangle, |1\rangle\}$. The server inserts $C_i$ into $S_i$ producing a new sequence $S_i^{'}$, and sends the new sequence ($S_i^{'}$) to $P_i$.

**Step (3)** Upon receiving $S_i^{'}$, every participant sends an acknowledgment to the server.

**Step (4)** In this step, the server announces the positions of $C_i$ and their measurement bases. Every $P_i$ measures $C_i$ based on the corresponding measurement bases and stores the results. Randomly, $P_i$ selects half of the qubits in $C_i$ and announces their measurement results to the server. The server, in turn, announces the initial states of the second half of $C_i$. Both the server and $P_i$ collaborate to compute the error rate. They end the protocol if the error rate is higher than a predefined value. Otherwise, $P_i$ discards $C_i$ from $S_i^{'}$ getting $S_i$ and continues to Step (5).

**Step (5)** After each $P_i$ gets the secure sequence $S_i$, they start to perform the next sub-steps.

a) *Encoding phase*. $P_i$ encodes the secret information ($K_i^{'}$) onto $S_i$ by applying the unitary operation $I = |0\rangle\langle 0| + |1\rangle\langle 1|$ when the classical bit $K_i^{'}$ is 0, and the unitary operation $U = |0\rangle\langle 1| - |1\rangle\langle 0|$ if the classical bit $K_i^{'}$ is 1 as indicated in Table 1.

b) *Detecting the external attack phase*. For detecting external eavesdroppers, $P_i$ generates a sequence of random single qubits ($C_{pi}$) as in Steps (1) and (2), which are used as decoy qubits to check the existence of eavesdroppers in the quantum channel between $P_i$ and $P_{i+1}$ (note, the symbol + in "$i+1$" represents the additional mod $n$. $P_i$ inserts $C_{pi}$ into $S_i$ producing a new sequence $S_{i\mapsto i+1}$, and sends the new sequence ($S_{i\mapsto i+1}$) to $P_{i+1}$. As in Step (4), $P_i$ and $P_{i+1}$ share the information of $C_{pi}$ and collaborate to compute the error rate. $P_i$ and $P_{i+1}$ end the protocol if the error rate is higher than a predefined value. Otherwise, $P_{i+1}$ discards $C_{pi}$ from $S_{i\mapsto i+1}$ obtaining $S_i$ and continues to the next process.

c) *Detecting the internal attack phase*. Upon confirming that the communication between $P_i$ and $P_{i+1}$ is secure against external attackers, the server randomly selects $l$ single-qubits as decoy qubits from $S_{i\mapsto i+1}$, by announcing their positions, and asks $P_i$ to publicly announce the unitary operations that were applied to the $l$ qubits. Subsequently, the server announces the measurement bases of the $l$ qubits to $P_{i+1}$. $P_{i+1}$ measures the $l$ qubits using the corresponding measurement bases. Based on the measurement results, the measurement bases, and the applied unitary operations, $P_{i+1}$ can judge whether the $l$ qubits are genuine or not. If not, $P_{i+1}$ ends the protocol. Otherwise, the participants do the following:

   i) $P_{i+1}$ discards the $l$ qubits from $S_{i\mapsto i+1}$ that were selected by the server;
   ii) The server also discards the corresponding $l$ qubits from $S_i$;
   iii) Every $P_i$ discards the corresponding classical bits from their private keys $K_i^{'}$.

d) As in Step (5.a), $P_{i+1}$ encodes the secret information ($K_{i+1}^{'}$) onto $S_i$ and inserts some random decoy states ($C_{pi+1}$) into $S_{i\mapsto i+1}$ producing $S_{i\mapsto i+2}$. Afterwards, $P_{i+1}$ sends $S_{i\mapsto i+2}$ to $P_{i+2}$.

e) Upon $P_{i+2}$ receiving $S_{i\mapsto i+2}$, $P_{i+1}$ and $P_{i+2}$ collaborate to check the security of communication by performing Step (5.a–5.d).

f) $P_{i+2}$ encodes her or his information and sends the new sequences to the next participants. This process continues until $P_i$ receives the secure quantum message ($S_{i\mapsto i-1}$) from $P_{i-1}$; here, the symbol "–" in "$i-1$" represents the subtraction mod $n$.

**Step (6):** When all $P_i$ s receive $S_{i\mapsto i-1}$, they discard the decoy qubits getting $S_i$. Hence, each participant loses $nl$ classical bits from $K_i^{'}$ getting $K_i$ with length $m$. After that, they send an acknowledgment to the server, and the server announces the measurement bases of $S_i$ to all the $P_i$ s. Finally, every $P_i$ uses the corresponding measurement bases to measure $S_i$ obtaining $K$, where $K = K_1 \oplus K_2 \oplus \cdots \oplus K_n$.

## Illustration of the proposed protocol

For simplicity, suppose we have three participants $P_1$, $P_2$, and $P_3$ and they want to generate a shared secret key $K = K_1 \oplus K_2 \oplus K_3$ with length $m$ (e.g., $m = 3$). $P_1$, $P_2$, and $P_3$ have three private keys $K_1^{'}$, $K_2^{'}$, and $K_3^{'}$, respectively, with length $m + nl$, e.g., $m + nl = 3 + (3 * 3) = 12$; here $nl$ is the number of decoy states for checking the security of all quantum channels in one complete circle, and for the $n$ circle it will be $m + nl$. Here, there are three complete circles for three participants, and the number of decoy qubits for checking the security of all quantum channels is $n * nl = 9l$. Also, we assume that, $K_1^{'} = 000001101101$, $K_2^{'} = 111011101000$, and $K_3^{'} = 110011010110$.

The server generates a sequence of quantum states contains 12 random states (e.g., $S_1 = |0\rangle, |0\rangle, |0\rangle, |1\rangle|0\rangle, |0\rangle, |1\rangle|0\rangle, |1\rangle, |-\rangle, |+\rangle, |-\rangle$) for the first circle and sends it to $P_1$. $P_1$ checks the security of the transmission with the server as in Step (4). Based on her/his private data ($K_1^{'}$), $P_1$ applies the unitary operations $\{I, I, I, I, I, U, U, I, U, U, I, U\}$ to $S_1$ getting $S_{1 \mapsto 2} = |0\rangle, |0\rangle, |0\rangle, |1\rangle|0\rangle, -|1\rangle, |0\rangle|0\rangle, |0\rangle, -|+\rangle, |+\rangle, -|+\rangle$. To secure the communication, $P_1$ inserts some decoy qubits into $S_{1 \mapsto 2}$ and sends $S_{1 \mapsto 2}$ to $P_2$. Subsequently, $P_2$ performs Step (5.b) to detect the external attack.

As in Step (5.c), the server chooses random $l$ states (e.g., $l = 1$) from $S_1$ and announce the position of $l$ (e.g., the position of last state in $S_1$) to $P_1$ and $P_2$. The server asks $P_1$ to announce the unitary operation that was applied to $l$, and asks $P_2$ to announce the measurement result of the corresponding states in $S_{1 \mapsto 2}$ (i.e., $-|+\rangle$), respectively. Based on the announced information ($|-\rangle, U, -|+\rangle$), the server can judge whether $P_2$ has received genuine information or not.

Later, the server and $P_2$ discard the last sequence from $S_1$ and $S_{1 \mapsto 2}$ getting newly updated sequences $S_1 = (|0\rangle, |0\rangle, |0\rangle, |1\rangle|0\rangle, |0\rangle, |1\rangle|0\rangle, |1\rangle, |-\rangle, |+\rangle)$ and $S_{1 \mapsto 2} = (|0\rangle, |0\rangle, |0\rangle, |1\rangle|0\rangle, -|1\rangle, |0\rangle|0\rangle, |0\rangle, -|+\rangle, |+\rangle)$, respectively. Also, all participants update their private keys by discarding the corresponding classical bits. The updated private keys of $P_1$, $P_2$, and $P_3$ become $K_1^{'} = 00000110110$, $K_2^{'} = 11101110100$, and $K_3^{'} = 11001101011$, respectively. They also consume two quantum states (e.g., the last two states) for checking the quantum channel between ($P_2$ and $P_3$) and ($P_3$ and $P_1$).

The updated private keys after completing one circle are as follows: $K_1^{'} = 000001101$, $K_2^{'} = 111011101$, and $K_3^{'} = 110011010$. And the updated private keys after completing the three circles are as follows: $K_1^{'} = 000$, $K_2^{'} = 111$, and $K_3^{'} = 110$. Now, $|K| = |K_1| = |K_2| = |K_3| = \left|K_1^{'}\right| = \left|K_2^{'}\right| = \left|K_2^{'}\right|$. Finally, each participant can get the final key $K = K_1 \oplus K_2 \oplus K_3 = 000 \oplus 111 \oplus 110 = 001$. Note that for simplicity, we assumed that the server frequently chooses the last qubit for checking the security of communication; but the selected positions should be completely random.

**Applying the proposed model to SCWZ's protocol.** Taking SCWZ's protocol[19] as an example, we show in this section how to address the vulnerability of CT-MQKA protocols to the collusive attack.

In SCWZ's protocol[19], there are $n$ participants and each participant $P_i$ ($i = 1, 2, \ldots, n$) has an *m-bit* key ($K_i$). All participants want to fairly generate a shared secret key ($K = K_1 \oplus K_2 \oplus \cdots \oplus K_n$). Also, there is a server that generates $n$ sequences of random single-photons. Each sequence $S_i$ contains $m$ random single-photons. The server generates additional $n$ sequences of random single photons ($C_i$), which are used as decoy photons to check the existence of eavesdroppers.

Based on our proposed model, SCWZ's protocol should be modified as follows.

1- Each participant ($P_i$) should prepare the length of her/his private keys ($K_i$) to be $m + nl$.
2- The length of the quantum sequences generated by the server should also be $m + nl$.
3- As in Step (5.b), $P_i$ should generate a sequence of random single-qubits ($C_{pi}$) to check the security of the quantum channel between the sender ($P_i$) and receiver ($P_{i+1}$).
4- To detect the collusive attack, the server randomly selects $l$ single-qubits from the $m + nl$ single-qubits and uses them as decoy qubits to check the security of quantum channels between every two participants, as proposed in Step (5.c).
5- All participants update their keys by discarding the classical bits corresponding to the single qubits that were used as decoy qubits.

**The security analysis.** This section presents detailed security analyses for both external eavesdropping and internal attacks.

**External attack.** In the proposed protocol, the decoy qubit technique is used to prevent external eavesdroppers from attacking the protocol. To achieve that, a sequence of single decoy qubits is randomly selected from the states $\{|+\rangle, |-\rangle, |0\rangle, |1\rangle\}$ and randomly inserted into the secret message. The eavesdropper (Eve) cannot distinguish between the decoy-states and secret message states. Eve may try to entangle a secret message state with an auxiliary quantum state ($|\epsilon\rangle$) by applying a unitary operation ($\boldsymbol{U_\epsilon}$) as follows:

$$U_\epsilon|0\rangle|\epsilon\rangle = \alpha_1|0\rangle|\epsilon_{00}\rangle + a_2|1\rangle|\epsilon_{01}\rangle, \quad (5)$$

$$U_\epsilon|1\rangle|\epsilon\rangle = \alpha_1|0\rangle|\epsilon_{10}\rangle + a_2|1\rangle|\epsilon_{11}\rangle, \quad (6)$$

$$U_\epsilon|+\rangle|\epsilon\rangle = \frac{1}{2}[|+\rangle(\alpha_1|\epsilon_{00}\rangle + \alpha_2|\epsilon_{01}\rangle + \alpha_3|\epsilon_{10}\rangle + \alpha_4|\epsilon_{11}\rangle) + |-\rangle(\alpha_1|\epsilon_{00}\rangle - \alpha_2|\epsilon_{01}\rangle + \alpha_3|\epsilon_{10}\rangle - \alpha_4|\epsilon_{11}\rangle)], \tag{7}$$

$$U_\epsilon|-\rangle|\epsilon\rangle = \frac{1}{2}[|+\rangle(\alpha_1|\epsilon_{00}\rangle + \alpha_2|\epsilon_{01}\rangle - \alpha_3|\epsilon_{10}\rangle - \alpha_4|\epsilon_{11}\rangle) + |-\rangle(\alpha_1|\epsilon_{00}\rangle - \alpha_2|\epsilon_{01}\rangle - \alpha_3|\epsilon_{10}\rangle + \alpha_4|\epsilon_{11}\rangle)]. \tag{8}$$

In (5) and (6), $|\alpha_1|^2 + |a_2|^2 = 1$ and $|\alpha_3|^2 + |\alpha_4|^2 = 1$. Also, $|\epsilon_{00}\rangle, |\epsilon_{01}\rangle, |\epsilon_{10}\rangle$, and $|\epsilon_{11}\rangle$ are four ancilla states decided by Eve. To pass the external eavesdropping detection phase, Eve sets $\alpha_2 = \alpha_3 = 0$, if the targeted quantum state is $|0\rangle$ or $|1\rangle$, and $(\alpha_1|e_{00}\rangle + \alpha_2|e_{01}\rangle - \alpha_3|e_{10}\rangle - \alpha_4|e_{11}\rangle) = (\alpha_1|e_{00}\rangle - \alpha_2|e_{01}\rangle + \alpha_3|e_{10}\rangle - \alpha_4|e_{11}\rangle) = 0$, if the targeted quantum state is $|+\rangle$ or $|-\rangle$. But these malicious procedures cannot help Eve to extract any useful information from the private inputs. For example, if Eve sets $\alpha_2 = \alpha_3 = 0$, she gets $|\alpha_1|^2 = |\alpha_4|^2 = 1$, which means that $\alpha_1|\epsilon_{00}\rangle = \alpha_4|\epsilon_{11}\rangle$. So, Eve cannot reveal private inputs. Besides, the proposed CT-MQKA protocol is not open to the Trojan horse attack since all information is sent in a one-way manner[30,31].

**Internal attack.** In the proposed model, internal attacks may be divided into three types of attacks: (1) Server's attack; (2) Participant's attack; (3) Collusive attack. Participant's attack is similar to server's attack in this work. Therefore, we only discuss here the server's attack and collusive attack.

*Server's attack.* In this work, we assume that the server is semi-honest. That is, it faithfully executes the operations delegated by participants and does not collude with other participants to steal sensitive information, but may try to get the information of secret keys. Participants employ the decoy photon method to secure the communications between every two participants. Hence, the server must adopt one of the external attack strategies if it wants to get sensitive information. However, we show in the "External attack" section that the proposed model is secure against external attacks. Accordingly, the malicious server may resort to guessing the required information or generate the final key as follows:

a) Passing the security check. In Step (1), the server sends $S_i$ to $P_i$ as an initial quantum sequence for generating the final key. In Step (5), $P_i$ uses $S_i$ to encode her/his private data and inserts some decoy qubits for security check before sending them to $P_{i+1}$. To successfully pass the security check, the server must correctly guess the measurement bases of the decoy qubits and guess the initial bases to correctly resend $P_i$'s qubits to $P_{i+1}$ without been caught. The probability of correctly guessing a measurement basis for each qubit is 50%, and the probability of correctly guessing an initial basis is also 50%. Therefore, the probability $(pr)$ of passing the eavesdropping check is as follows:

$$pr = pr_1 \times pr_2 \times \cdots \times pr_n = \left(\frac{1}{2} \times \frac{1}{2}\right)^{nl} \times \left(\frac{1}{2} \times \frac{1}{2}\right)^{nl} \times \cdots \times \left(\frac{1}{2} \times \frac{1}{2}\right)^{nl} = \left(\frac{1}{2} \times \frac{1}{2}\right)^{nl}. \tag{9}$$

here, $pr_i$ $(i = 1, 2, \ldots, n)$ is the probability of correctly guessing the $i$th sequence of decoy qubits, and $l$ is the length of each decoy qubit sequence.

b) Guessing participants' private keys. Since $K = K_1 \oplus K_2 \oplus \cdots \oplus K_n$, the server needs to correctly guess all participants' private keys to get $K$. The probability $(pr)$ of correctly guessing the final key $K$ is as follows:

$$pr = pr_1 \times pr_2 \times \cdots \times pr_n = \left(\frac{1}{2}\right)^{l} \times \left(\frac{1}{2}\right)^{l} \times \ldots \times \left(\frac{1}{2}\right)^{l} = \left(\frac{1}{2}\right)^{nl}. \tag{10}$$

here, $pr_i$ $(i = 1, 2, \ldots, n)$ is the probability of correctly guessing $K_i$, and $l$ is the length of $K_i$.

c) Guessing the final key $(K)$. The server may try to directly guess the final key $(K)$. In that case, the probability $(pr)$ is as follows:

$$pr = \left(\frac{1}{2}\right)^{l}, \text{ where } l \text{ is the length of } K. \tag{11}$$

In Eqs. (9–11), if $l$ is large enough the probability of guessing the final key or required information is close to zero or negligible.

*Collusive attack.* A collusive attack is the most powerful internal attack in which two or more dishonest participants collude together to extract sensitive information or generate the final key alone without revealing their malicious behaviour. In this section, we show that the proposed model is immune to collusive attacks, such that any group of dishonest participants trying to perform a collusive attack (including the two attack strategies mentioned in the section *The insecurity of existing CT-MQKA protocols*) will be detected immediately.

Basically, dishonest participants rely on two important processes to successfully achieve the collusive attack; (1) sharing information about the carrier quantum states that will be used to encode the private data and generate the final key, (2) deceiving the honest participants to deduce their private data by sending forged data. In our protocol, a semi-honest server is used to prevent dishonest participants from sending forged data to the honest ones. The server generates $S_i$ with decoy qubits in Step (1) and sends it to $P_i$ in Step (2). In Step (5.b), $P_i$ sends her/his encoded sequence to $P_{i+1}$. To prevent the collusive attack, in Step (5. c), the server participates in checking the security of transmission to make sure that $P_i$ does not send forged qubits to $P_{i+1}$ by randomly selecting some qubits and asking the participants to divulge the related information. Accordingly, the protocol guarantees that the honest participant has received genuine data, and the dishonest participants cannot obtain useful information to generate the final key alone or steal the private inputs of honest participants.

Moreover, if the dishonest participants try to adopt guessing strategies they will be detected with high probability as indicated in Eqs. (9–11). Thus, we can say that the proposed model is secure against internal attacks.

## Conclusion

In this work, we showed that most of the existing circular-type multiparty quantum key agreement protocols are insecure against a specific type of collusive attack. We analyzed the security of a recently proposed circular-type multiparty quantum key agreement protocol to demonstrate the vulnerability of such protocols. Then, we proposed a general secure quantum key agreement model to avoid the different types of collusive attacks. We showed that the proposed protocol could generate the final key correctly and that the proposed protocol is secure against all known collusive attack strategies.

## References

1. Diffie, W. & Hellman, M. New directions in cryptography. *IEEE Trans. Inf. Theory* **22**, 644–654 (1976).
2. Ingemarsson, I., Tang, D. & Wong, C. A conference key distribution system. *IEEE Trans. Inf. Theory* **28**, 714–720 (1982).
3. Pieprzyk, J. & Li, C.-H. Multiparty key agreement protocols. *IEE Proc. Comput. Digital Tech.* **147**, 229–236 (2000).
4. Bernstein, D. J. & Lange, T. Post-quantum cryptography. *Nature* **549**, 188–194 (2017).
5. Alagic, G. *et al. Status Report on the Second Round of the NIST Post-Quantum Cryptography Standardization Process* (US Department of Commerce, 2020).
6. Bennet, C. & Brassard, G. in *Proc. of IEEE Int. Conf. on Comp., Syst. and Signal Proc., Bangalore, India, Dec. 10–12* (1984).
7. Abulkasim, H., Farouk, A., Hamad, S., Mashatan, A. & Ghose, S. Secure dynamic multiparty quantum private comparison. *Sci. Rep.* **9**, 1–16 (2019).
8. Abulkasim, H. *et al.* Improving the security of quantum key agreement protocols with single photon in both polarization and spatial-mode degrees of freedom. *Quant. Inf. Process.* **17**, 316 (2018).
9. Wu, W., Cai, Q., Wu, S. & Zhang, H. Cryptanalysis of He's quantum private comparison protocol and a new protocol. *Int. J. Quant. Inf.* **17**, 1950026 (2019).
10. Qi, R. *et al.* Implementation and security analysis of practical quantum secure direct communication. *Light Sci. Appl.* **8**, 1–8 (2019).
11. Li, L. & Li, Z. A verifiable multiparty quantum key agreement based on bivariate polynomial. *Inf. Sci.* **521**, 343–349 (2020).
12. Abulkasim, H. & Alotaibi, A. Improvement on 'Multiparty Quantum Key Agreement with Four-Qubit Symmetric W State'. *Int. J. Theor. Phys.* **58**, 4235–4240 (2019).
13. Shi, R.-H. & Zhang, M. Privacy-preserving quantum sealed-bid auction based on grover's search algorithm. *Sci. Rep.* **9**, 1–10 (2019).
14. Bao, N. & Halpern, N. Y. Quantum voting and violation of Arrow's impossibility theorem. *Phys. Rev. A* **95**, 062306 (2017).
15. Chowdhury, A. *et al.* Quantum signature of a squeezed mechanical oscillator. *Phys. Rev. Lett.* **124**, 023601 (2020).
16. Abulkasim, H. *et al.* Improved dynamic multi-party quantum private comparison for next-generation mobile network. *IEEE Access* **7**, 17917–17926 (2019).
17. Zhou, N., Zeng, G. & Xiong, J. Quantum key agreement protocol. *Electron. Lett.* **40**, 1149–1150 (2004).
18. Liu, B., Xiao, D., Jia, H.-Y. & Liu, R.-Z. Collusive attacks to "circle-type" multi-party quantum key agreement protocols. *Quantum Inf. Process.* **15**, 2113–2124 (2016).
19. Sun, Z., Cheng, R., Wu, C. & Zhang, C. New fair multiparty quantum key agreement secure against collusive attacks. *Sci. Rep.* **9**, 1–8 (2019).
20. Sun, Z., Wu, C., Zheng, S. & Zhang, C. Efficient multiparty quantum key agreement with a single d-level quantum system secure against collusive attack. *IEEE Access* **7**, 102377–102385 (2019).
21. Liu, H.-N., Liang, X.-Q., Jiang, D.-H., Zhang, Y.-H. & Xu, G.-B. Multi-party quantum key agreement protocol with bell states and single particles. *Int. J. Theor. Phys.* **58**, 1659–1666 (2019).
22. Liu, W.-J., Chen, Z.-Y., Ji, S., Wang, H.-B. & Zhang, J. Multi-party semi-quantum key agreement with delegating quantum computation. *Int. J. Theor. Phys.* **56**, 3164–3174 (2017).
23. Shi, R.-H. & Zhong, H. Multi-party quantum key agreement with bell states and bell measurements. *Quantum Inf. Process.* **12**, 921–932 (2013).
24. Wang, L. & Ma, W. Quantum key agreement protocols with single photon in both polarization and spatial-mode degrees of freedom. *Quantum Inf. Process.* **16**, 130 (2017).
25. Liu, H.-N., Liang, X.-Q., Jiang, D.-H., Xu, G.-B. & Zheng, W.-M. Multi-party quantum key agreement with four-qubit cluster states. *Quantum Inf. Process.* **18**, 242 (2019).
26. Huang, W. *et al.* Efficient multiparty quantum key agreement with collective detection. *Sci. Rep.* **7**, 1–9 (2017).
27. Zhou, Y.-H., Zhang, J., Shi, W.-M., Yang, Y.-G. & Wang, M.-F. Continuous-variable multiparty quantum key agreement based on third party. *Mod. Phys. Lett. B* **34**, 2050083 (2020).
28. Cao, H. & Ma, W. Multi-party traveling-mode quantum key agreement protocols immune to collusive attack. *Quantum Inf. Process.* **17**, 219 (2018).
29. Huang, W.-C., Yang, Y.-K., Jiang, D., Gao, C.-H. & Chen, L.-J. Designing secure quantum key agreement protocols against dishonest participants. *Int. J. Theor. Phys.* **58**, 4093–4104 (2019).
30. Li, X.-H., Deng, F.-G. & Zhou, H.-Y. Improving the security of secure direct communication based on the secret transmitting order of particles. *Phys. Rev. A* **74**, 054302 (2006).
31. Deng, F.-G., Li, X.-H., Zhou, H.-Y. & Zhang, Z.-J. Improving the security of multiparty quantum secret sharing against Trojan horse attack. *Phys. Rev. A* **72**, 044302 (2005).

## Acknowledgements

This research was funded by the Natural Sciences and Engineering Research Council of Canada and NXM Labs Inc. NXM's autonomous security technology enables devices, including connected vehicles, to communicate securely with each other and their surroundings without human intervention while leveraging data at the edge to provide business intelligence and insights. NXM ensures data privacy and integrity by using a novel blockchain-based architecture which enables rapid and regulatory-compliant data monetization. Ryerson University is in the "Dish With One Spoon Territory". The Dish With One Spoon is a treaty between the Anishinaabe, Mississaugas and Haudenosaunee that bound them to share the territory and protect the land. Subsequent Indigenous Nations and peoples, Europeans and all newcomers, have been invited into this treaty in the spirit of peace, friendship and respect. Wilfrid Laurier University is located on the traditional territory of the Neutral, Anishnawbe and Haudenosaunee peoples. We thank them for allowing us to conduct research on their land.

## Author contributions

H.A. designed the scheme. H.A., A.M. and S.G. did security analysis and comparisons. All authors contributed to the writing and discussion of the paper.

## Competing interests

The authors declare no competing interests.

## Additional information

**Correspondence** and requests for materials should be addressed to H.A.

**Reprints and permissions information** is available at www.nature.com/reprints.

**Publisher's note** Springer Nature remains neutral with regard to jurisdictional claims in published maps and institutional affiliations.